\documentclass[aps,prl,superscriptaddress,showpacs,floatfix,twocolumn]{revtex4}

\usepackage{graphicx}	

\begin{document}


\title{$J/\psi$ Production and Nuclear Effects for $d+Au$ and $p+p$
       Collisions at $\sqrt{s_{NN}}$ = 200~GeV }

\newcommand{\abilene}{Abilene Christian University, Abilene, TX 79699, USA}
\newcommand{\acadsin}{Institute of Physics, Academia Sinica, Taipei 11529, Taiwan}
\newcommand{\banaras}{Department of Physics, Banaras Hindu University, Varanasi 221005, India}
\newcommand{\barc}{Bhabha Atomic Research Centre, Bombay 400 085, India}
\newcommand{\bnl}{Brookhaven National Laboratory, Upton, NY 11973-5000, USA}
\newcommand{\caucr}{University of California - Riverside, Riverside, CA 92521, USA}
\newcommand{\ciae}{China Institute of Atomic Energy (CIAE), Beijing, People's Republic of China}
\newcommand{\cns}{Center for Nuclear Study, Graduate School of Science, University of Tokyo, 7-3-1 Hongo, Bunkyo, Tokyo 113-0033, Japan}
\newcommand{\colorado}{University of Colorado, Boulder, CO 80309, USA}
\newcommand{\columbia}{Columbia University, New York, NY 10027 and Nevis Laboratories, Irvington, NY 10533, USA}
\newcommand{\dapnia}{Dapnia, CEA Saclay, F-91191, Gif-sur-Yvette, France}
\newcommand{\debrecen}{Debrecen University, H-4010 Debrecen, Egyetem t{\'e}r 1, Hungary}
\newcommand{\elte}{ELTE, E{\"o}tv{\"o}s Lor{\'a}nd University, H - 1117 Budapest, P{\'a}zm{\'a}ny P. s. 1/A, Hungary}
\newcommand{\fsu}{Florida State University, Tallahassee, FL 32306, USA}
\newcommand{\gsu}{Georgia State University, Atlanta, GA 30303, USA}
\newcommand{\hiroshima}{Hiroshima University, Kagamiyama, Higashi-Hiroshima 739-8526, Japan}
\newcommand{\ihepprot}{IHEP Protvino, State Research Center of Russian Federation, Institute for High Energy Physics, Protvino, 142281, Russia}
\newcommand{\illuiuc}{University of Illinois at Urbana-Champaign, Urbana, IL 61801, USA}
\newcommand{\isu}{Iowa State University, Ames, IA 50011, USA}
\newcommand{\jinrdubna}{Joint Institute for Nuclear Research, 141980 Dubna, Moscow Region, Russia}
\newcommand{\kek}{KEK, High Energy Accelerator Research Organization, Tsukuba, Ibaraki 305-0801, Japan}
\newcommand{\kfki}{KFKI Research Institute for Particle and Nuclear Physics of the Hungarian Academy of Sciences (MTA KFKI RMKI), H-1525 Budapest 114, POBox 49, Budapest, Hungary}
\newcommand{\korea}{Korea University, Seoul, 136-701, Korea}
\newcommand{\kurchatov}{Russian Research Center ``Kurchatov Institute", Moscow, Russia}
\newcommand{\kyoto}{Kyoto University, Kyoto 606-8502, Japan}
\newcommand{\labllr}{Laboratoire Leprince-Ringuet, Ecole Polytechnique, CNRS-IN2P3, Route de Saclay, F-91128, Palaiseau, France}
\newcommand{\lawllnl}{Lawrence Livermore National Laboratory, Livermore, CA 94550, USA}
\newcommand{\losalamos}{Los Alamos National Laboratory, Los Alamos, NM 87545, USA}
\newcommand{\lpc}{LPC, Universit{\'e} Blaise Pascal, CNRS-IN2P3, Clermont-Fd, 63177 Aubiere Cedex, France}
\newcommand{\lund}{Department of Physics, Lund University, Box 118, SE-221 00 Lund, Sweden}
\newcommand{\muenster}{Institut f\"ur Kernphysik, University of Muenster, D-48149 Muenster, Germany}
\newcommand{\myongji}{Myongji University, Yongin, Kyonggido 449-728, Korea}
\newcommand{\nagasaki}{Nagasaki Institute of Applied Science, Nagasaki-shi, Nagasaki 851-0193, Japan}
\newcommand{\newmex}{University of New Mexico, Albuquerque, NM 87131, USA }
\newcommand{\nmsu}{New Mexico State University, Las Cruces, NM 88003, USA}
\newcommand{\ornl}{Oak Ridge National Laboratory, Oak Ridge, TN 37831, USA}
\newcommand{\orsay}{IPN-Orsay, Universite Paris Sud, CNRS-IN2P3, BP1, F-91406, Orsay, France}
\newcommand{\peking}{Peking University, Beijing, People's Republic of China}
\newcommand{\pnpi}{PNPI, Petersburg Nuclear Physics Institute, Gatchina, Leningrad region, 188300, Russia}
\newcommand{\riken}{RIKEN (The Institute of Physical and Chemical Research), Wako, Saitama 351-0198, JAPAN}
\newcommand{\rikjrbrc}{RIKEN BNL Research Center, Brookhaven National Laboratory, Upton, NY 11973-5000, USA}
\newcommand{\saopaulo}{Universidade de S{\~a}o Paulo, Instituto de F\'{\i}sica, Caixa Postal 66318, S{\~a}o Paulo CEP05315-970, Brazil}
\newcommand{\seoulnat}{System Electronics Laboratory, Seoul National University, Seoul, South Korea}
\newcommand{\stonybrkc}{Chemistry Department, Stony Brook University, Stony Brook, SUNY, NY 11794-3400, USA}
\newcommand{\stonycrkp}{Department of Physics and Astronomy, Stony Brook University, SUNY, Stony Brook, NY 11794, USA}
\newcommand{\subatech}{SUBATECH (Ecole des Mines de Nantes, CNRS-IN2P3, Universit{\'e} de Nantes) BP 20722 - 44307, Nantes, France}
\newcommand{\tenn}{University of Tennessee, Knoxville, TN 37996, USA}
\newcommand{\titech}{Department of Physics, Tokyo Institute of Technology, Oh-okayama, Meguro, Tokyo 152-8551, Japan}
\newcommand{\tsukuba}{Institute of Physics, University of Tsukuba, Tsukuba, Ibaraki 305, Japan}
\newcommand{\vandy}{Vanderbilt University, Nashville, TN 37235, USA}
\newcommand{\waseda}{Waseda University, Advanced Research Institute for Science and Engineering, 17 Kikui-cho, Shinjuku-ku, Tokyo 162-0044, Japan}
\newcommand{\weizmann}{Weizmann Institute, Rehovot 76100, Israel}
\newcommand{\yonsei}{Yonsei University, IPAP, Seoul 120-749, Korea}
\newcommand{\deceased}{\dagger}
\affiliation{\abilene}
\affiliation{\acadsin}
\affiliation{\banaras}
\affiliation{\barc}
\affiliation{\bnl}
\affiliation{\caucr}
\affiliation{\ciae}
\affiliation{\cns}
\affiliation{\colorado}
\affiliation{\columbia}
\affiliation{\dapnia}
\affiliation{\debrecen}
\affiliation{\elte}
\affiliation{\fsu}
\affiliation{\gsu}
\affiliation{\hiroshima}
\affiliation{\ihepprot}
\affiliation{\illuiuc}
\affiliation{\isu}
\affiliation{\jinrdubna}
\affiliation{\kek}
\affiliation{\kfki}
\affiliation{\korea}
\affiliation{\kurchatov}
\affiliation{\kyoto}
\affiliation{\labllr}
\affiliation{\lawllnl}
\affiliation{\losalamos}
\affiliation{\lpc}
\affiliation{\lund}
\affiliation{\muenster}
\affiliation{\myongji}
\affiliation{\nagasaki}
\affiliation{\newmex}
\affiliation{\nmsu}
\affiliation{\ornl}
\affiliation{\orsay}
\affiliation{\peking}
\affiliation{\pnpi}
\affiliation{\riken}
\affiliation{\rikjrbrc}
\affiliation{\saopaulo}
\affiliation{\seoulnat}
\affiliation{\stonybrkc}
\affiliation{\stonycrkp}
\affiliation{\subatech}
\affiliation{\tenn}
\affiliation{\titech}
\affiliation{\tsukuba}
\affiliation{\vandy}
\affiliation{\waseda}
\affiliation{\weizmann}
\affiliation{\yonsei}
\author{S.S.~Adler}	\affiliation{\bnl}
\author{S.~Afanasiev}	\affiliation{\jinrdubna}
\author{C.~Aidala}	\affiliation{\columbia}
\author{N.N.~Ajitanand}	\affiliation{\stonybrkc}
\author{Y.~Akiba}	\affiliation{\kek} \affiliation{\riken}
\author{A.~Al-Jamel}	\affiliation{\nmsu}
\author{J.~Alexander}	\affiliation{\stonybrkc}
\author{K.~Aoki}	\affiliation{\kyoto}
\author{L.~Aphecetche}	\affiliation{\subatech}
\author{R.~Armendariz}	\affiliation{\nmsu}
% ...
\collaboration{PHENIX Collaboration} \noaffiliation

\date{\today}

\begin{abstract}
$J/\psi$ production in $d+Au$ and $p+p$ collisions at 
$\sqrt{s_{NN}} = 200$ GeV has been measured by the PHENIX experiment at
rapidities $-2.2 < y < +2.4$.  The cross sections and nuclear dependence
of $J/\psi$ production versus rapidity, transverse momentum, and
centrality are obtained and compared to lower energy $p+A$ results and to
theoretical models. The observed nuclear dependence in $d+Au$ collisions
is found to be modest, suggesting that the absorption in the final state
is weak and the shadowing of the gluon distributions is small and
consistent with Dokshitzer-Gribov-Lipatov-Altarelli-Parisi-based
parameterizations that fit deep-inelastic
scattering and Drell-Yan data at lower energies.
\end{abstract}



\maketitle


$J/\psi$ production in hadron collisions, since it proceeds predominantly through
diagrams involving gluons (e.g. gluon fusion)~\cite{Bedjidian:2003gd}, is a
sensitive probe of the gluon structure function in the nucleon and its
modification in nuclei. It is also a leading signal for the creation of hot-dense
matter in heavy-ion collisions~\cite{Matsui:1986dk}. Shadowing of partons (quarks
or gluons) in nuclei is a depletion of their population at small momentum fraction
of the nucleon, $x$, compared to that in a free nucleon, with a corresponding enhancement
at moderate $x$ (anti-shadowing).
In $\sqrt{s_{NN}} = 200$ GeV deuteron-gold
($d+Au$) collisions at the Relativistic Heavy Ion Collider (RHIC), for positive
(deuteron direction) rapidities, gluons are probed that lie well into the
shadowing region with momentum fractions in $Au$, $x_2 \sim 3 \times 10^{-3}$.
Models of gluon shadowing predict suppressions of $J/\psi$ production in nuclei
that differ by as much as a factor of
three~\cite{Eskola:2001gt,Frankfurt:1998ym,Kopeliovich:2001ee}. Recent theoretical
developments, e.g. the Color Glass Condensate model~\cite{McLerran:1993ni},
suggest that at very low $x$ non-linear gluon saturation effects
become important and cause substantial modifications of the gluon densities.

The connection of the observed $J/\psi$ suppression to the modified gluon distribution
in nuclei can be clouded by the absorption of the final-state
$c\bar{c}$~\cite{Vogt:1999dw} which depends on the poorly known production
mechanism~\cite{Bedjidian:2003gd} and by the energy loss of the intial-state gluon
- although the latter is thought to be small at RHIC energies~\cite{Kopeliovich:2001ee}.
This connection is also distorted by the fact that approximately a third of the
$J/\psi^{\ \prime}$s come from decays of higher-mass resonances~\cite{Antoniazzi:1992iv}.

Here we present measurements made by the PHENIX experiment at RHIC for the
production of $J/\psi^{\ \prime}$s in $\sqrt{s_{NN}} = 200$ GeV $d+Au$ and
proton-proton ($p+p$) collisions. These data provide the first measurement of the
nuclear dependence of $J/\psi$ production at this energy, a much higher energy
than previous $p+A$ measurements from
fixed-target experiments at $\sqrt{s_{NN}} \lesssim 40$
GeV~\cite{Alde:1990wa,Leitch:1999ea,Alessandro:2003pi,Spengler:2004gr,Badier:1983dg}.
Although our measurements are for $d+A$, the nuclear effects on the $J/\psi$ in 
deuterium were found to be small at lower energies~\cite{Leitch:2004aa}.
Besides the shadowing region at small $x$, these data also probe larger gluon
momentum fractions (at negative rapidity) nearer the rest frame of the residual
nucleus. Finally, these measurements also serve as a baseline for the upcoming
results from the high-luminosity $Au+Au$ and $Cu+Cu$ runs and must be understood
in order to look for effects beyond what is expected from cold nuclear matter.

The measurements described here are similar to earlier ones with PHENIX~\cite{Adler:2003NIM} for
$p+p$~\cite{Adler:2003qs} and $Au+Au$~\cite{Adler:2003rc} collisions, but with a
second muon spectrometer added and higher luminosity. The two muon spectrometers
are especially valuable for asymmetric collisions such as $d+Au$ where
simultaneous measurements at positive ($1.2 < y < 2.4$) and negative ($-2.2 < y <
-1.2$) rapidities, along with central $(|y| \le 0.35)$ rapidity from $e^+e^-$, are
then available. Electrons in the central arms are identified by matching charged
particle tracks to clusters in an electromagnetic calorimeter (EMC) and to rings
in a ring imaging \v{C}erenkov (RICH) detector. Muons are identified by their
detection in Iarocci tubes after their penetration through 8 to 11
interaction lengths of copper and steel absorber.

The data used in this analysis were recorded in 2003 using a trigger that required
hits in each of the two beam-beam counters located at negative and positive
rapidity ($3 < |\eta| < 3.9$). In addition, for the di-muons at least two tracks
in the muon identifier of appropriate absorber depth were required, while for the
di-electrons a one-track trigger with a signal above threshold in the EMC with a
matching hit in the RICH was required. After quality and vertex cuts, the samples
for the three arms correspond to integrated luminosities from 180 to 250~nb$^{-1}$
($p+p$) and 1.4 to 1.7~nb$^{-1}$ ($d+Au$).

For the di-muons the $J/\psi$ yield is obtained after subtraction of the
combinatoric background using like-sign muon pairs ($2 \sqrt{N_{++} N_{--}}$)
and by fitting the resulting mass peak with a Gaussian plus an exponential to
represent the small remaining continuum background underneath the peak. A variety
of continuum shapes were checked for each fit in order to establish the
uncertainty due to the low-statistics background. For the di-electrons the
combinatoric background was subtracted using the sum of like-sign pairs and the
$J/\psi$ yield was taken as all remaining events in the mass range 2.6 to 3.6
GeV/$c^2$. A total ($p+p$ plus $d+Au$) of about 2100 and 500 $J/\psi^{\prime}$s
were obtained in the $\mu\mu$ and $ee$ channels, respectively.

The differential cross sections are calculated as,
\begin{equation} 
B_{ll} 
\frac{ d\sigma_{J/\psi} }{ dy } \,=\,
\frac{ N_{J/\psi} }
{ A \, \epsilon_{rec} \, \epsilon_{trig} \, \epsilon_{J/\psi}^{BBC} \,
( N_{evt}/{ (\sigma^{tot}_{MB} \cdot \epsilon_{MB}^{BBC}) } ) } \,
\frac{ 1 } { \Delta y } \,
\label{eq:dsigd_alt}
\end{equation}

\noindent and the nuclear modification factor, $R_{dA}$, is

\begin{equation} 
R_{dA} =\,
\frac{ d\sigma^{dAu}_{J/\psi}/dy }{ (2 \times 197) \times d\sigma^{pp}_{J/\psi}/dy } \,
\label{eq:rda_alt}
\end{equation}

In the above expressions $B_{ll}$ is the $J/\psi$ branching ratio to di-leptons,
$N_{J/\psi}$ is the measured $J/\psi$ yield, $A$ is the geometrical acceptance,
$\epsilon_{rec}$ is the di-lepton reconstruction efficiency, $\epsilon_{trig}$ is
the trigger efficiency, $N_{evt}$ is the number of min-bias triggers 
sampled,
$\sigma^{tot}_{MB}$ is the total minimum-bias (MB) cross section, $\Delta y$ is the
rapidity bin width, and $\epsilon_{MB}^{BBC}$ and $\epsilon_{J/\psi}^{BBC}$ are the
beam-beam trigger efficiencies for min-bias and $J/\psi$ events respectively. The
factor of $2 \times 197$ causes $R_{dA}$ to be one if the $d+A$ cross section is
just additive from $p+p$, i.e. if there are no nuclear modifications.

For $p+p$ we use the cross section for our beam-beam trigger, ${\sigma^{tot}_{MB}}(pp) 
\epsilon_{MB}^{BBC}(pp) = 23.0 \pm 2.2$~mb; and the efficiency for events with a
$J/\psi$, $\epsilon_{J/\psi}^{BBC}(pp) = 0.79 \pm 0.02$. For
$d+Au$ collisions we use a beam-beam trigger cross section of
$\sigma^{tot}_{MB}(dAu) \epsilon_{MB}^{BBC}(dAu) = 1.99 \pm 0.10$~b
from our measurement~\cite{Swhite:2005swhite} using photo-dissociation of the deuteron
as a reference~\cite{Klein:2002pi}, which is consistent with our calculated Glauber result of
$1.92\pm0.18$~b. For the $J/\psi$ we use $\epsilon_{J/\psi}^{BBC}(dAu) = 
0.94 \pm 0.02$~\cite{Adler:2004eh}.

For $d+Au$ collisions, the centrality of the collision can be characterized by
measuring the charge deposited in the beam-beam counter in the $Au$ beam
direction~\cite{Adler:2004eh}. An approximate number of nucleon+nucleon collisions
$\langle N_{coll} \rangle$ can be obtained through a Glauber calculation that
relates this $\langle N_{coll} \rangle$ to the observed charge. In this case
$R_{dA}(N_{coll})$ is calculated as,

\begin{equation} 
R_{dA}({\langle N_{coll} \rangle}) =\,
\frac{ N^{dAu}_{inv}({\langle N_{coll} \rangle}) }{ \langle N_{coll} \rangle \times N^{pp}_{inv} } \,
\label{eq:rdacent_alt}
\end{equation}

\noindent where the invariant yield $N_{inv}$ is,

\begin{equation} 
N_{inv}({\langle N_{coll} \rangle}) \,=\,
\frac{ N_{J/\psi} C_{bias}({\langle N_{coll} \rangle}) }{ A \, \epsilon_{rec} \, \epsilon_{trig} [N_{evt} (\Delta w / w) ] } \,
\label{eq:ninv_alt}
\end{equation}

\noindent with $\langle N_{coll} \rangle$ being the average number of binary
collisions for a particular $d+Au$ centrality bin and $N_{evt} (\Delta w /
w)$ the number of $d+Au$ min-bias triggers sampled that lie in this fraction,
$\Delta w$, of the total minimum-bias centrality range, $w$.
This prescription is equivalent to that of Eq. 1,2 for minimum bias.
For $p+p$ collisions $\Delta w / w$ is one. $C_{bias}=\epsilon_{MB}^{BBC}/\epsilon_{J/\psi}^{BBC}$ is a
correction for the smaller trigger efficiency in minimum-bias events compared to
those with a $J/\psi$. For $d+Au$, $C_{bias}$ depends on ${\langle N_{coll}
\rangle}$ and takes into account the effect of the underlying event multiplicity
on both the trigger efficiency and the centrality measurement~\cite{Adler:2004eh}.
Its variation with centrality is up to 7\% from unity.


For the electron analysis, $A \, \epsilon_{rec} \,$ and $\epsilon_{trig}$ were
determined using a {\sc Geant}~\cite{GEANT:W5013} simulation of the central arms
and a trigger response software emulation~\cite{Adler:2003qs}. $\epsilon_{rec}$
was confirmed by studying pairs identified as photon conversions in the data. The
systematic uncertainty of 10.4\% is dominated by run-to-run efficiencies (5\%),
yield extraction (5\%), and the occupancy dependence of the efficiency (4.4\%).


For the muon arms, $A \, \epsilon_{rec} \, \epsilon_{trig}$ was determined within
each rapidity and $p_T$ bin, using a {\sc Geant} simulation with $J/\psi$ events
generated by {\sc Pythia}~\cite{Sjostrand:2000wi}. The dominant systematic
uncertainties in our result are +6/$-$9\% from the muon identifier efficiency and up
to 10\% (for the most central negative rapidity $d+Au$ data) from the combinatoric
background.

 \begin{figure}[tbhp]
 \includegraphics[width=1.0\linewidth]{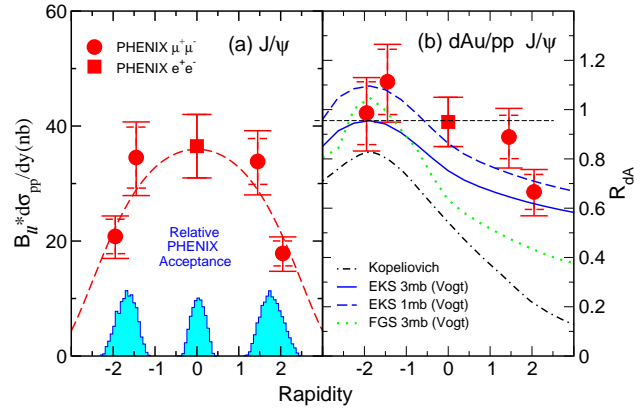}
 \caption{\label{fig:sigpp_rda} (color online)
(a) The 200 GeV $J/\psi$ $p+p$ differential cross section times di-lepton
branching ratio versus rapidity (10\% overall normalization uncertainty is not
included). (b) The minimum bias $R_{dA}$ versus rapidity (12\% overall
normalization uncertainty not included).  For both panels the dashed error bars
represent systematic uncertainties relevant for comparing the two rapidity bins in
each muon arm, while the solid error bars represent the overall uncertainties
relevant for comparing points at negative, central, or positive rapidity. The
curve in (a) represents a fit as described in the text while the curves in (b) are
theoretical calculations~\cite{Klein:2003dj,Vogt:2004dh,Kopeliovich:2001ee} as
described in the text.}
\end{figure}


Figure~\ref{fig:sigpp_rda}~(a) shows the measured $p+p$ differential cross
section times branching ratio versus rapidity with a di-electron point at
mid-rapidity and two di-muon points at negative and positive rapidities. A fit to
a shape generated with {\sc Pythia} is performed and, using a di-lepton branching
ratio of 5.9\%~\cite{Eidelman:2004pdb}, gives a total cross section
$\sigma_{pp}^{J/\psi} = 2.61 \pm 0.20 ({\rm fit}) \pm 0.26 ({\rm abs}) ~\mu$b.
Variations in the parton distribution functions and models used to determine the
shape are negligible compared to the fit errors. This result is smaller by about
two sigma than our previous lower statistics result~\cite{Adler:2003qs}.


Figure~\ref{fig:sigpp_rda}~(b) shows the nuclear modification factor 
$R_{dA}$ (Eq. 2) 
versus rapidity, where a value of one would correspond to no nuclear
modification.  While this ratio is consistent with unity at negative rapidity, it
is significantly lower at the most positive rapidity where gluons are expected to
be shadowed in a heavy nucleus.  Theoretical
predictions~\cite{Klein:2003dj,Vogt:2004dh,Kopeliovich:2001ee} that include the
effects of absorption and shadowing are shown for comparison in
Fig.~\ref{fig:sigpp_rda}~(b).  The data favor a relatively modest 
shadowing in
agreement with the parametrization of Eskola-Kolhinen-Salgado (EKS) based 
on a leading-twist Dokshitzer-Gribov-Lipatov-Altarelli-Parisi-evolved 
parametrization of nuclear deep inelastic scattering
and Drell-Yan data at lower energies~\cite{Eskola:2001gt}, rather than the stronger gluon shadowing of
Kopeliovich~\cite{ Kopeliovich:2001ee} or 
Frankfurt-Guzey-Strikman (FGS)~\cite{Frankfurt:1998ym} based on
models involving coherence for a $q\bar{q}$ dipole in the nucleus.
The $c{\bar c}$ absorption cross section is not well determined by 
our data, but is probably nearer
to 1 mb and is certainly smaller than the $4.1 \pm 0.4$ mb found at lower energy~\cite{na50abs}.

\begin{figure}[tbhp]
 \includegraphics[width=1.0\linewidth]{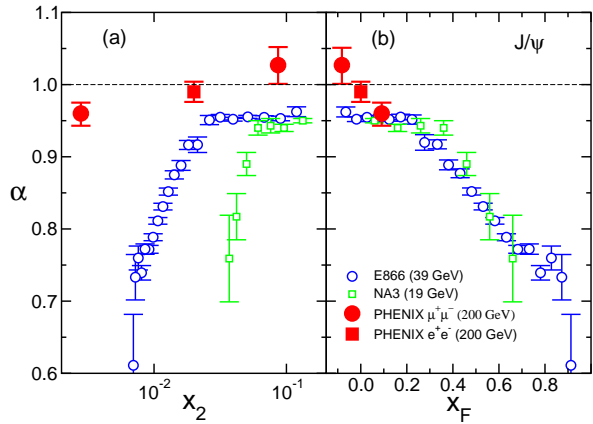}
 \caption{\label{fig:alpha_x2xf} (color online) $\alpha$ versus
(a) $\alpha$ versus $p_T$ compared to lower energy 
measurements shown for three different $x_F$ ranges.
An additional overall uncertainty of 0.02 in our $\alpha$ values is not 
shown. }
\end{figure}

Lower energy $p+A$ measurements showed that $J/\psi$ suppression did not follow a
universal behavior vs $x_2$~\cite{x2def}, the momentum fraction in the heavy nucleus,
as would be expected if the suppression was dominated by shadowing~\cite{Leitch:1999ea}.
As shown in Fig.~\ref{fig:alpha_x2xf}~(a), our data confirm this $x_2$ 
scaling violation
with the addition of a smaller $x_2$ point, but with our small 
range in $x_F$ have little
to add to the approximate $x_F$ scaling observed in these lower energy measurements
[see Fig. ~\ref{fig:alpha_x2xf}~(b)].
Here $\alpha$ is defined by $\sigma_{dA} = \sigma_{pp} \times (2A)^{\alpha}$ and
$x_F = x_1 - x_2$, where $x_1$ is the momentum fraction of the gluon in the deuteron.
This $x_F$ scaling may be caused by energy loss of the gluon
in the intial state~\cite{Vogt:1999dw},
or by an energy conservation effect (``Sudakov suppression")~\cite{Kopeliovich:2005aa} which causes a
universal suppression that increases with $x_F$ for production of $J/\psi$'s and other hadrons.
It is also similar to that observed for more positive rapidity hadron
production in the $d+Au$ ``limiting fragmentation'' region~\cite{Back:2003hx}.


\begin{figure*}[thbp]
 \includegraphics[width=0.75\linewidth]{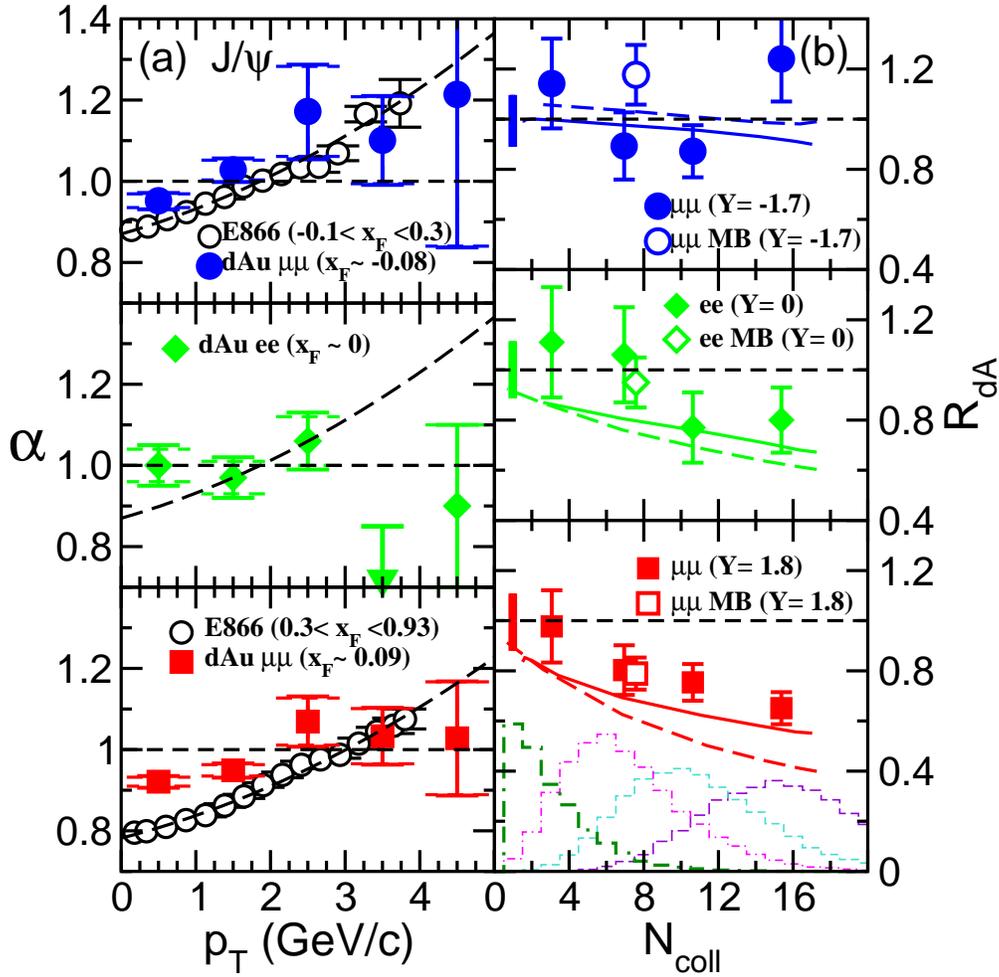}
 \caption{\label{fig:pt_ncoll} (color online)
(a) $\alpha$ versus $p_T$ compared
to lower energy measurements, shown for three different $x_F$ ranges.
The error bars have the same meaning as in Fig.~\ref{fig:sigpp_rda}.
An additional 0.02 overall uncertainty is not shown.
The dashed curves are simple fits~\cite{Leitch:1999ea} to the lower energy results.
(b) Nuclear modification factor versus
centrality as given by the number of nucleon+nucleon collisions shown for three
different rapidity ranges, compared to theoretical
calculations\cite{Klein:2003dj,Vogt:2004dh} including final-state absorption and
EKS (solid) or FGS (dashed) shadowing. The bars at the low end of each plot
represent the systematic errors between different rapidity ranges. An additional
12\% global error bar is not shown. The histograms at the bottom of the lower
panel indicate the distribution of the number of collisions for each of the four
centrality bins.
}
\end{figure*}

Invariant cross sections versus transverse momentum, $(d^2\sigma/dyd
p_T)/(2\pi p_T)$, have been fit to the form $A \times
(1+(p_T/B)^2)^{-6}$~\cite{Yoh:1978id}. Average $p^2_T$ values resulting
from these fits are $4.28 \pm 0.31$, $3.03 \pm 0.40$ and $3.63 \pm 0.25$
GeV/$c^2$ for $d+Au$ collisions at negative, zero, and positive $x_F$,
respectively;  compared with $2.51 \pm 0.21$ and $4.31 \pm 0.85$ GeV/$c^2$
for negative/positive and zero $x_F$ $p+p$ collisions, respectively. The
observed $p_T$ broadening is shown in Fig.~\ref{fig:pt_ncoll}~(a). For
negative $x_F$ it is consistent with that of the lower energy
($\sqrt{s_{NN}}=39$ GeV) measurements from
E866/NuSea~\cite{Leitch:1999ea}, but may be flatter at positive $x_F$. At
zero $x_F$ no $p_T$ broadening is seen within errors.


$R_{dA}$ (Eq. 3) is
shown in Fig.~\ref{fig:pt_ncoll}~(b) for four centrality classes and for 
minimum bias
collisions. This classification into centrality bins for these results can only be
approximate, as indicated by the overlapping histograms of $N_{coll}$.  At positive
rapidity (small $x_2$, or the shadowing region), a weak drop for more central
collisions is observed, while no significant centrality dependence is seen for
negative rapidity or for central rapidity. The theoretical curves on
Fig.~\ref{fig:pt_ncoll}~(b) correspond to different amounts of density 
dependent
shadowing and anti-shadowing~\cite{Klein:2003dj,Vogt:2004dh} and also include
absorption. They are consistent with our data except at positive rapidity where
the EKS shadowing curve is closest to our results, although slightly lower perhaps
due to the amount of absorption that is included.


In summary, during the RHIC 2003 run, the PHENIX experiment measured
nuclear effects on $J/\psi$ production for $d+Au$ collisions at
$\sqrt{s_{NN}}$ = 200~GeV. Increasing suppression for larger rapidity
(smaller $x_2$) and for more central collisions (higher nuclear
densities sampled) both are consistent with models containing a small
amount of impact-parameter dependent shadowing and with weak absorption.
Theoretical calculations which include EKS shadowing seem most consistent
with the data. However comparisons with other measurements at lower
energies show that shadowing cannot be the dominant effect, at least not
for the lower energy measurements. We also see some transverse momentum
broadening which is consistent with that seen at lower energy. Higher
luminosity $d+Au$ running in the future yielding higher numbers of
$J/\psi^{\ \prime}$s will be necessary to quantify these nuclear effects
and to more clearly distinguish between various theoretical models of
shadowing, absorption and other cold nuclear matter effects.


We thank the staff of the Collider-Accelerator and Physics
Departments at BNL for their vital contributions.  We acknowledge
support from the Department of Energy and NSF (U.S.A.),
MEXT and JSPS (Japan), CNPq and FAPESP (Brazil), NSFC (China),
IN2P3/CNRS, CEA, and ARMINES (France),
BMBF, DAAD, and AvH (Germany),
OTKA (Hungary), DAE and DST (India), ISF (Israel),
KRF and CHEP (Korea), RMIST, RAS, and RMAE (Russia),
VR and KAW (Sweden), U.S. CRDF for the FSU,
US-Hungarian NSF-OTKA-MTA, and US-Israel BSF.


\begin{thebibliography}{25}
\expandafter\ifx\csname natexlab\endcsname\relax\def\natexlab#1{#1}\fi
\expandafter\ifx\csname bibnamefont\endcsname\relax
  \def\bibnamefont#1{#1}\fi
\expandafter\ifx\csname bibfnamefont\endcsname\relax
  \def\bibfnamefont#1{#1}\fi
\expandafter\ifx\csname citenamefont\endcsname\relax
  \def\citenamefont#1{#1}\fi
\expandafter\ifx\csname url\endcsname\relax
  \def\url#1{\texttt{#1}}\fi
\expandafter\ifx\csname urlprefix\endcsname\relax\def\urlprefix{URL }\fi
\providecommand{\bibinfo}[2]{#2}
\providecommand{\eprint}[2][]{\url{#2}}

\bibitem[{\citenamefont{Bedjidian et~al.}(2003)}]{Bedjidian:2003gd}
\bibinfo{author}{\bibfnamefont{M.}~\bibnamefont{Bedjidian}}
  \bibnamefont{et~al.} \eprint{hep-ph/0311048} (\bibinfo{year}{2003}).

\bibitem[{\citenamefont{Matsui and Satz}(1986)}]{Matsui:1986dk}
\bibinfo{author}{\bibfnamefont{T.}~\bibnamefont{Matsui}} \bibnamefont{and}
  \bibinfo{author}{\bibfnamefont{H.}~\bibnamefont{Satz}},
  \bibinfo{journal}{Phys. Lett.} \textbf{\bibinfo{volume}{B178}},
  \bibinfo{pages}{416} (\bibinfo{year}{1986}).

\bibitem[{\citenamefont{Eskola et~al.}(2001)\citenamefont{Eskola, Kolhinen, and
  Vogt}}]{Eskola:2001gt}
\bibinfo{author}{\bibfnamefont{K.~J.} \bibnamefont{Eskola}},
  \bibinfo{author}{\bibfnamefont{V.~J.} \bibnamefont{Kolhinen}},
  \bibnamefont{and} \bibinfo{author}{\bibfnamefont{R.}~\bibnamefont{Vogt}},
  \bibinfo{journal}{Nucl. Phys.} \textbf{\bibinfo{volume}{A696}},
  \bibinfo{pages}{729} (\bibinfo{year}{2001}).

\bibitem[{\citenamefont{Frankfurt and Strikman}(1999)}]{Frankfurt:1998ym}
\bibinfo{author}
{\bibfnamefont{L.}~\bibnamefont{Frankfurt}}, 
{\bibfnamefont{V.}~\bibnamefont{Guzey}}, 
\bibnamefont{and}
  \bibinfo{author}{\bibfnamefont{M.}~\bibnamefont{Strikman}},
  \bibinfo{journal}{Eur. Phys. J.} \textbf{\bibinfo{volume}{A5}},
  \bibinfo{pages}{293} (\bibinfo{year}{1999}).

\bibitem[{\citenamefont{Kopeliovich et~al.}(2001)\citenamefont{Kopeliovich,
  Tarasov, and H{\"u}fner}}]{Kopeliovich:2001ee}
\bibinfo{author}{\bibfnamefont{B.}~\bibnamefont{Kopeliovich}},
  \bibinfo{author}{\bibfnamefont{A.}~\bibnamefont{Tarasov}}, \bibnamefont{and}
  \bibinfo{author}{\bibfnamefont{J.}~\bibnamefont{H{\"u}fner}},
  \bibinfo{journal}{Nucl. Phys.} \textbf{\bibinfo{volume}{A696}},
  \bibinfo{pages}{669} (\bibinfo{year}{2001}).

\bibitem[{\citenamefont{McLerran and Venugopalan}(1994)}]{McLerran:1993ni}
\bibinfo{author}{\bibfnamefont{L.} \bibnamefont{McLerran}} \bibnamefont{and}
  \bibinfo{author}{\bibfnamefont{R.}~\bibnamefont{Venugopalan}},
  \bibinfo{journal}{Phys. Rev. D} \textbf{\bibinfo{volume}{49}},
  \bibinfo{pages}{2233} (\bibinfo{year}{1994}).

\bibitem[{\citenamefont{Vogt}(2000)}]{Vogt:1999dw}
\bibinfo{author}{\bibfnamefont{R.}~\bibnamefont{Vogt}}, \bibinfo{journal}{Phys.
  Rev. C} \textbf{\bibinfo{volume}{61}}, \bibinfo{pages}{035203}
  (\bibinfo{year}{2000}).

\bibitem[{\citenamefont{Antoniazzi et~al.}(1993)}]{Antoniazzi:1992iv}
\bibinfo{author}{\bibfnamefont{L.}~\bibnamefont{Antoniazzi}}
  \bibnamefont{et~al.} 
\bibinfo{journal}{Phys. Rev. Lett.} 
\textbf{\bibinfo{volume}{70}}, \bibinfo{pages}{383}
  (\bibinfo{year}{1993}).

\bibitem[{\citenamefont{Alde et~al.}(1991)}]{Alde:1990wa}
\bibinfo{author}{\bibfnamefont{D.~M.} \bibnamefont{Alde}} \bibnamefont{et~al.},
  \bibinfo{journal}{Phys. Rev. Lett.} \textbf{\bibinfo{volume}{66}},
  \bibinfo{pages}{133} (\bibinfo{year}{1991}).

\bibitem[{\citenamefont{Leitch et~al.}(2000)}]{Leitch:1999ea}
\bibinfo{author}{\bibfnamefont{M.~J.} \bibnamefont{Leitch}}
  \bibnamefont{et~al.}, 
  \bibinfo{journal}{Phys. Rev. Lett.} \textbf{\bibinfo{volume}{84}},
  \bibinfo{pages}{3256} (\bibinfo{year}{2000}).

\bibitem[{\citenamefont{Alessandro et~al.}(2003)}]{Alessandro:2003pi}
\bibinfo{author}{\bibfnamefont{B.}~\bibnamefont{Alessandro}}
  \bibnamefont{et~al.},
\bibinfo{journal}{Phys.
  Lett.} \textbf{\bibinfo{volume}{B553}}, \bibinfo{pages}{167}
  (\bibinfo{year}{2003}).

\bibitem[{\citenamefont{Spengler}(2004)}]{Spengler:2004gr}
\bibinfo{author}{\bibfnamefont{J.}~\bibnamefont{Spengler}}
\bibinfo{journal}{J. Phys.}
  \textbf{\bibinfo{volume}{G30}}, \bibinfo{pages}{S871} 
  (\bibinfo{year}{2004}).

\bibitem[{\citenamefont{Badier et~al.}(1983)}]{Badier:1983dg}
\bibinfo{author}{\bibfnamefont{J.}~\bibnamefont{Badier}} 
\bibnamefont{et~al.},
\bibinfo{journal}{Z. Phys.}
  \textbf{\bibinfo{volume}{C20}}, \bibinfo{pages}{101} (\bibinfo{year}{1983}).

\bibitem[{\citenamefont{Leitch}(2004)}]{Leitch:2004aa}
\bibinfo{author}{\bibfnamefont{M.~J.}~\bibnamefont{Leitch}}, 
\bibinfo{journal}{Eur. Phys. J A}
  \textbf{\bibinfo{volume}{S19}}, \bibinfo{pages}{129} 
(\bibinfo{year}{2004}).

\bibitem[{\citenamefont{Adler et~al.}(2003)}]{Adler:2003NIM}
\bibinfo{author}{\bibfnamefont{S.~S.} \bibnamefont{Adler}} \bibnamefont{et~al.},
  \bibinfo{journal}{ Nucl. Instrum. Meth.} \textbf{\bibinfo{volume}{A499}},
  \bibinfo{pages}{469} (\bibinfo{year}{2003}).

\bibitem[{\citenamefont{Adler et~al.}(2004{\natexlab{a}})}]{Adler:2003qs}
\bibinfo{author}{\bibfnamefont{S.~S.} \bibnamefont{Adler}} 
\bibnamefont{et~al.},
\bibinfo{journal}{Phys. Rev. Lett.}
  \textbf{\bibinfo{volume}{92}}, \bibinfo{pages}{051802}
  (\bibinfo{year}{2004}{\natexlab{a}}).

\bibitem[{\citenamefont{Adler et~al.}(2004{\natexlab{b}})}]{Adler:2003rc}
\bibinfo{author}{\bibfnamefont{S.~S.} \bibnamefont{Adler}} 
\bibnamefont{et~al.},
\bibinfo{journal}{Phys. Rev. C}
  \textbf{\bibinfo{volume}{69}}, \bibinfo{pages}{014901}
  (\bibinfo{year}{2004}{\natexlab{b}}).

\bibitem[{\citenamefont{White}(2005)}]{Swhite:2005swhite}
\bibinfo{author}{\bibfnamefont{S.~N.} \bibnamefont{White}},
  \bibinfo{journal}{{\em Proceedings of the XIIIth International Workshop DIS'05,
  Madison, WI, 2005}, edited by W.H. Smith 
[AIP Conf. Proc., \textbf{792}, No. 1, 527, (2005)]}.  

\bibitem[{\citenamefont{Klein and Vogt}(2003{\natexlab{a}})}]{Klein:2002pi}
\bibinfo{author}{\bibfnamefont{S.~R.} \bibnamefont{Klein}} \bibnamefont{and}
  \bibinfo{author}{\bibfnamefont{R.}~\bibnamefont{Vogt}},
  \bibinfo{journal}{Phys. Rev. C} \textbf{\bibinfo{volume}{67}},
  \bibinfo{pages}{047901} (\bibinfo{year}{2003}{\natexlab{a}}).

\bibitem[{\citenamefont{Adler et~al.}(2005)}]{Adler:2004eh}
\bibinfo{author}{\bibfnamefont{S.~S.} \bibnamefont{Adler}} 
\bibnamefont{et~al.},
\bibinfo{journal}{Phys. Rev. Lett.}
  \textbf{\bibinfo{volume}{94}}, \bibinfo{pages}{082302}
  (\bibinfo{year}{2005}).

\bibitem[{GEA(1993)}]{GEANT:W5013}
\bibinfo{author}{\bibfnamefont{F.} \bibnamefont{Carminati}}
\bibnamefont{et~al.},
 \emph{\bibinfo{title}{``GEANT: Detector Description and Simulation Tool"}}, 
\bibinfo{journal}
{CERN Program Library Long Writeup No. W5013, 1993 (unpublished)}.

\bibitem[{\citenamefont{Sjostrand et~al.}(2001)}]{Sjostrand:2000wi}
\bibinfo{author}{\bibfnamefont{T.}~\bibnamefont{Sjostrand}}
  \bibnamefont{et~al.}, \bibinfo{journal}{Comput. Phys. Commun.}
  \textbf{\bibinfo{volume}{135}}, \bibinfo{pages}{238} (\bibinfo{year}{2001}).

\bibitem[{\citenamefont{Eidelman et~al.}(2004)}]{Eidelman:2004pdb}
\bibinfo{author}{\bibfnamefont{S.}~\bibnamefont{Eidelman}} 
\bibnamefont{et~al.},
\bibinfo{journal}{Phys. Lett.} 
\textbf{\bibinfo{volume}{B592}}, \bibinfo{pages}{1}
  (\bibinfo{year}{2004}).

\bibitem[{\citenamefont{Klein and Vogt}(2003{\natexlab{b}})}]{Klein:2003dj}
\bibinfo{author}{\bibfnamefont{S.~R.} \bibnamefont{Klein}} \bibnamefont{and}
  \bibinfo{author}{\bibfnamefont{R.}~\bibnamefont{Vogt}},
  \bibinfo{journal}{Phys. Rev. Lett.} \textbf{\bibinfo{volume}{91}},
  \bibinfo{pages}{142301} (\bibinfo{year}{2003}{\natexlab{b}}).

\bibitem[{\citenamefont{Vogt}(2005)}]{Vogt:2004dh}
\bibinfo{author}{\bibfnamefont{R.}~\bibnamefont{Vogt}}, \bibinfo{journal}{Phys.
  Rev. C} \textbf{\bibinfo{volume}{71}}, \bibinfo{pages}{054902}
  (\bibinfo{year}{2005}).

\bibitem[{\citenamefont{Borges et~al.}(2004)}]{na50abs}
\bibinfo{author}{\bibfnamefont{G.}~\bibnamefont{Borges}},
\bibnamefont{et~al.} \eprint{hep-ex/0505065} (\bibinfo{year}{2004}).

\bibitem[{\citenamefont{}()}]{x2def}
\bibinfo{title}{$x_{2} = 0.5(-x_F +\sqrt{x_F^2+4M_{J/\psi}^2/s_{NN}})$}.

\bibitem[{\citenamefont{Kopeliovich}(2005)}]{Kopeliovich:2005aa}
\bibinfo{author}{\bibfnamefont{B.~Z.}~\bibnamefont{Kopeliovich}}, \bibnamefont{et~al.} \eprint{hep-ph/0501260}
  (\bibinfo{year}{2005}).

\bibitem[{\citenamefont{Back et~al.}(2004)}]{Back:2003hx}
\bibinfo{author}{\bibfnamefont{B.~B.} \bibnamefont{Back}} 
\bibnamefont{et~al.},
\bibinfo{journal}{Phys. Rev. Lett.}
  \textbf{\bibinfo{volume}{93}}, \bibinfo{pages}{082301}
  (\bibinfo{year}{2004}).

\bibitem[{\citenamefont{Yoh et~al.}(1978)}]{Yoh:1978id}
\bibinfo{author}{\bibfnamefont{J.~K.} \bibnamefont{Yoh}} \bibnamefont{et~al.},
  \bibinfo{journal}{Phys. Rev. Lett.} \textbf{\bibinfo{volume}{41}},
  \bibinfo{pages}{684} (\bibinfo{year}{1978}).

\end{thebibliography}


\end{document}